\documentclass{styles/svproc}
\usepackage{url}
\usepackage{hyperref}

\usepackage{graphicx}
\usepackage{amsmath}
\usepackage{mathtools}
\usepackage{multicol}
\usepackage{amssymb}
\usepackage{colortbl} 
\usepackage{xcolor}
\usepackage{cleveref}
\usepackage{tcolorbox}
\usepackage{tikz}     

\begin{document}
\mainmatter              
\title{Interpretable Data-Driven Ship Dynamics Model: Enhancing Physics-Based Motion Prediction with Parameter Optimization}
\titlerunning{Interpretable Data-Driven Ship Dynamics Model}  
%

\author{ Christos Papandreou \and  Michail Mathioudakis \and
 Theodoros Stouraitis \and  Petros Iatropoulos \and  Antonios Nikitakis \and
Stavros Paschalakis \and Konstantinos Kyriakopoulos}
\authorrunning{Papandreou et al.} 
%
\tocauthor{Papandreou Christos, Mathioudakis Michail, Stouraitis Theodoros, Iatropoulos Petros, Nikitakis Antonios,
Stavros Paschalakis and Konstantinos Kyriakopoulos}
\institute{DeepSea Technologies, Greece\\
\email{c.papandreou@deepsea.ai},\\ 
home page:
\texttt{https://www.deepsea.ai/}
}

\maketitle              

\begin{abstract} 
The deployment of autonomous navigation systems on ships necessitates accurate motion prediction models tailored to individual vessels. Traditional physics-based models, while grounded in hydrodynamic principles, often fail to account for ship-specific behaviors in real-world conditions. Conversely, purely data-driven models offer specificity but lack interpretability and robustness in edge cases. This study proposes a data-driven physics-based model that integrates physics-based equations with data-driven parameter optimization, leveraging the strengths of both approaches to ensure interpretability and adaptability.
The model incorporates physics-based components such as 3-DoF dynamics, rudder, and propeller forces, while parameters such as resistance curve and rudder coefficients are optimized using synthetic data. By embedding domain knowledge into the parameter optimization process, the fitted model maintains physical consistency.  
Validation of the approach is realized with two container ships by comparing, both qualitatively and quantitatively, predictions against ground-truth trajectories. The results demonstrate significant improvements, in predictive accuracy and reliability, of the data-driven physics-based models over baseline physics-based models tuned with traditional marine engineering practices. The fitted models capture ship-specific behaviors in diverse conditions with their predictions being, 51.6\% (ship A) and 57.8\% (ship B) more accurate, 72.36\% (ship A) and 89.67\% (ship B) more consistent.

\end{abstract}

\section{Introduction}
Motion prediction of marine vessels is essential for safe and efficient navigation of autonomous ships. Thus, it needs to be (i) reliable to minimize risks and enable precise maneuvers, \textit{e.g.} in-port and collision avoidance, and (ii) accurate to facilitate informed decision-making, route optimization, and economical propulsion under the influence of dynamic environmental conditions such as currents, waves, and wind. 

Motion prediction is typically built upon physics-based models that leverage well-established hydrodynamic principles to be reliable. Yet, these models often struggle to capture ship-specific behaviors influenced by hull design, propulsion dynamics, and operational conditions. Physics-based models may lack the specificity needed for unique vessel conditions, leading to inaccurate predictions. Conversely, purely data-driven models can be tailored to individual ships to predict its motion accurately, but are often treated as black-boxs that are challenging to validate in unforeseen conditions, hence they lack interpretability and reliability.

Combining these two approaches into a data-driven physics-based model (grey-box) can mitigate these challenges, as a physics-based structure with data-driven parameter optimization can maintain interpretability while ensuring specificity. The objective of this paper is to present a data-driven physics-based model that integrates physics principles with data-driven adaptation methods. Specific parameters of the physics-based model that are key to its accuracy and reliability are identified and are marked as unknown. Using a constraint non-linear least squares (cNLS) optimization method~\cite{stephen2022convex} the values of these parameters are fitted to improve the prediction accuracy of the motion prediction.
The resulting model achieves superior predictive accuracy, and improved reliability while preserving the interpretability of ship's modeling aspects across diverse operating conditions. The key contributions of this work are: (i) development of a data-driven physics-based model that maintains physical consistency while incorporating vessel-specific behavioral patterns, (ii) implementation of an optimization framework that successfully adapts physics parameters to vessel-specific characteristics utilizing vessel data, and (iii) validation with two vessels across diverse operational conditions, with spatial coverage.


The rest of the paper is organized as follows: \cref{sec:related_Work} reviews  related work in ship motion modeling. \cref{sec:framework} describes our proposed framework, detailing its components and optimization. \cref{sec:results} presents the evaluation setup and results. Finally, \cref{sec:conclusion} concludes the study and, discusses future work.  \vspace{-2mm}

\section{Related Work}
\label{sec:related_Work}

\textbf{Physics-based (white-box) models}:
Research on maneuvering models began with~\cite{davidson1946turning}. Later,\cite{motora1959measurement},\cite{motora1955course}, and~\cite{nomoto1957steering} contributed to hydrodynamic aspects and stability principles. \cite{abkowitz1964ship} and~\cite{aastrom1976identification} used Taylor expansion for maneuvering models. The MMG model that is widely used in autonomous navigation was introduced by~\cite{ogawa1977mmg}, was improved by~\cite{yasukawa2015introduction} and has been applied by~\cite{li2013active} and~\cite{zhang2017ship}. Other models for low-speed hydrodynamic forces include the HD model~\cite{yasukawa2015introduction}, the CD model~\cite{yoshimura2012hydrodynamic}, and the TBL model~\cite{sutulo2015development}.

\textbf{Black box models}: Recently machine Learning approaches such as   artificial neural networks~\cite{moreira2003dynamic,rajesh2008system,zhang2013black,oskin2013neural,luo2016modeling,woo2018dynamic} and Gaussian processes~\cite{arizaramirez2018nonparametric,xue2020system} have been explored. \cite{SILVA2022103222} uses deep neural networks (DNN) for 6-DoF motion prediction in waves to demonstrate how DNNs can capture nonlinear dynamics between the ship state and wave conditions while maintaining good generalization.

\textbf{Grey box models}:
An other research avenue has focused on developing parameter estimation methods to combine traditional physics-based models with data-driven adaptation.
\cite{suyama2024parameter,MiyauchiCMAES} employed a gradient-free optimization method (namely CMA-ES) to tune parameters of an MMG model~\cite{yasukawa2015introduction} using sea-trial data of a real-scale ship. 
\cite{kanazawa2023bridging} proposed a framework that 
fits open parameters of an Abkowitz model~\cite{abkowitz1964ship} to data of the target ship while using a well-validated model of a similar ship to regularize the fitting process. In this way, the resulting predictions are on par with the validated model while they are tailored to the target ship's behavioral patterns. 
Although, improved prediction accuracy has been demonstrated, depending on the type of fitted parameters reliability to the model can be lost, if the parameters are not interpretable.
Unlike these approaches, the data-driven physics-based model introduced in this paper, which is based on the white-box model~\cite{mathioudakis2025realworldvalidationphysicsbasedship}, focuses on optimizing interpretable parameters, making the results more accessible and understandable to domain experts.   \vspace{-4mm}



\begin{figure}[t]
    \includegraphics[width=1\linewidth]{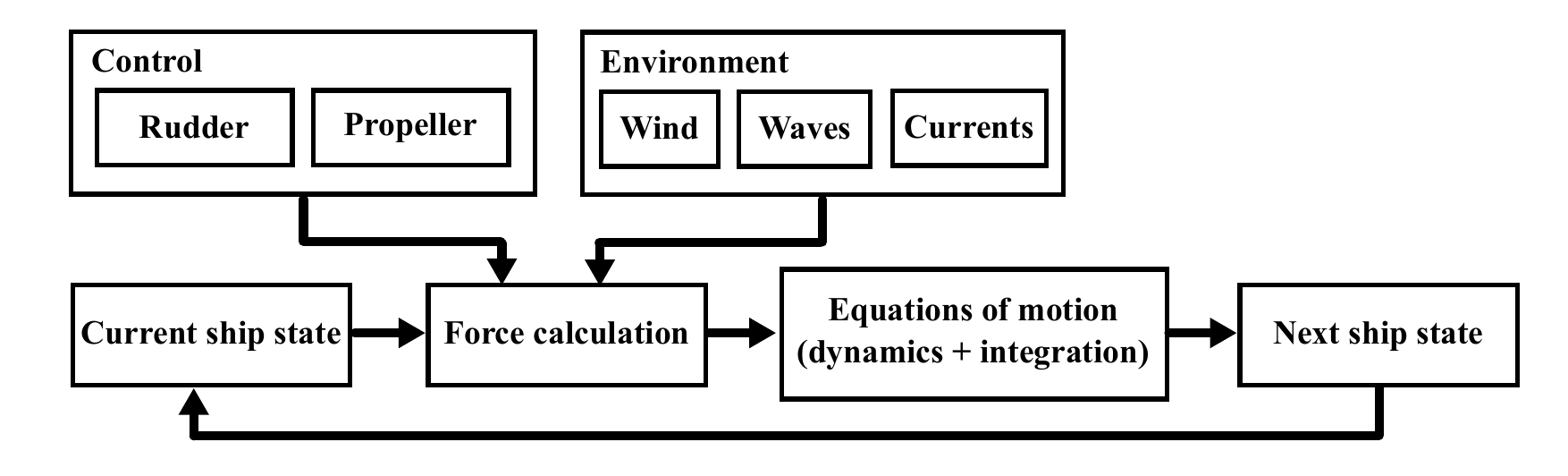} 
    \vspace{-8mm}
    \caption{Computational flow of a physics-based vessel motion prediction model, with the following key blocks: (a) ``Control": input commands of the rudder and the propeller, (b) ``Environment": environmental effects from the wind, waves, and sea currents, (c) ``Force calculation": computation of forced using the outputs of (a) and (b) along with hydrodynamic forces and the current ship state, and (d) ``Equations of motion": dynamics equations and integration over time to produce the next ship state.}
    \label{fig:simulation_system}
    \vspace{-3mm}
\end{figure}

\section{Prediction Model and Method}
\label{sec:framework}
In this section, first we briefly describe the motion prediction process, the primary aspects of the physics-based model, and the respective notation. Second, we introduce the open parameters of the physics-based model that are selected for fitting, and third we provide the details on the cNLS fitting method.  \vspace{-1mm}

\subsection{Motion Prediction and Model}
\label{subsec:model}

The state of the vessel has 3-DoF and it is denoted as $\mathbf{s} = [x, y, \psi, u, v, r, n, \delta]^T$, where $x$ and $y$ are the vessel cartesian coordinates, $\psi$ is the heading, $u$, $v$, $r$ are the surge, sway and angular velocities, $n$ is the propeller rpm and $\delta$ is the rudder angle. The $q$ inputs to the system are denoted with $\mathbf{u} \in \mathbb{R}^q$ and they comprise of: (i) control commands for the propeller $c_n$ and the rudder $c_\delta$, and (ii) environmental factors denoted with $e$, such as wind, waves, and currents. 

As shown in~\cref{fig:simulation_system}, given the current state of the ship $\mathbf{s}_k$, and a sequence of inputs $\mathbf{u}_k, \forall k \in \{0,..,K-1\}$, the motion prediction iteratively generates a trajectory $\xi = \{\mathbf{s}_0, ..., \mathbf{s}_{K}\}$ that is sequence of $K$ (number of knots) vessel states. The trajectory generation process is recursive and it involves at each timestep the calculation of: (a) the actuation forces from propeller and rudder, and (b) the environmental forces from wind, waves, and currents, and (c) the numerical integration (explicit Euler) of the motion equations from $\mathbf{s}_k$ to $\mathbf{s}_{k+1}$, as per~\cite{mathioudakis2025realworldvalidationphysicsbasedship}.

For the equations of motion (dynamics), a 3-DoF nonlinear ODE from~\cite{spyrou1996dynamic} is used. This is an HD model (see~\cref{sec:related_Work}) described by~\cref{eq:dynamics}.
\begin{equation}
    \label{eq:dynamics}
    \begin{split}
        (m - X_{\dot{u}})  \dot{u} +  (Y_{\dot{v}} -X_{vr} -m) vr + (Y_{\dot{r}} - m x_G) r^2 \\
        (m - Y_{\dot{v}}) \dot{v} + (m x_G -  Y_{\dot{r}})\dot{r} + (-Y_{v}U v + (mu - Y_rU)r - \hspace{1cm}\\ Y_{vv} v |v| - Y_{vr} v |r| - Y_{rr} r |r|) \\
        (m x_G - N_{\dot{v}}) \dot{v} + (I_z - N_{\dot{r}})\dot{r} - N_v U v 
         + (m x_G u - N_rU) r - \hspace{1cm}\\ 
         N_{rr} r |r| - N_{rrv} \frac{rrv}{U} - N_{vvr} \frac{vvr}{U} 
    \end{split}
    \begin{split}
        = \Sigma X \\
        \\
        = \Sigma Y  \\
        \\
        = \Sigma N 
    \end{split}
\end{equation} 
In~\cref{eq:dynamics}, m is the mass of the ship, $I_{zz}$ is the yaw moment of inertia and $x_G$ is the longitudinal distance of ship’s centre of gravity from the moving axes’ origin, $O_s$. $X$, $Y$ and $N$ terms with velocities $u$, $v$, $r$ and accelerations $\dot{u},\ \dot{v},\ \dot{r}$ subscripts are the maneuvering coefficients of the ship per dimension, which capture information regarding the hydrodynamic effects based on its geometric characteristics. In general, the left-hand-side (LHS) of~\eqref{eq:dynamics} incorporates terms that express the added mass, damping and restoring forces phenomena. Parameters on the LHS are assumed to be accurate as the hydrodynamic coefficients are calculated for each ship according to~\cite{Clarke1982TheAO} and~\cite{inoue1981hydrodynamic}. 

The right-hand-side (RHS) of~\cref{eq:dynamics} collects all the external forces and moments acting on the ship, which are expanded in~\cref{eq:forces}. 
\begin{equation}
\label{eq:forces}
    \begin{split}
        \Sigma X &= X_{n} + X_\delta + X_{e} + R(u) \\
        \Sigma Y &= Y_{n} + Y_\delta + Y_{e} \\
        \Sigma N &= N_{n} + N_\delta + N_{e}
    \end{split} 
\end{equation}
For each force dimension, the subscripts denote its source and $R(u) \in \mathbb{R}$ denotes the the resistance of the vessel in Newton. In this work, we study the effect of fitting key parameters on the RHS of~\cref{eq:dynamics}, hence parameters of~\cref{eq:forces}. 






\subsection{Key Parameters}
\label{subsec:open_parameters}
Considering that the hybrid model's adaptability and interpretability stems from carefully selecting the parameters to be fitted (see grey box models in~\cref{sec:related_Work}), we identified eleven key parameters. These parameters are chosen to enhance three pivotal aspects of ship behavior while maintaining physical significance. The three aspects of ship behavior follow.

\textbf{Propeller thrust}: For the propeller's thrust $X_n$ we use a fully defined propulsion model~\cite{mathioudakis2025realworldvalidationphysicsbasedship}, and we estimate the lateral force $Y_n$ generated by the propeller according to $ Y_{n} = p_{0} X_{n} $.
In this way, a single parameter relates the two dimensions of the thrust linearly, while the moment $N_n$ is calculated geometrically based on the propeller's position and $Y_n$.
Parameter $p_{0}$ models course instabilities and captures the lateral force produced by the propeller.
 
\textbf{Vessel resistance}: The resistance is a $3_{rd}$ order polynomial curve, described as $ R(u) = p_1 u + p_2 u^2 +p_3 u^3 $,
characterizes the ship's hull resistance through water. Obtaining a resistance curve 
that is appropriately balanced with the trust $X_n$, enhances the model's calculation accuracy of the ship's surge speed $u$. 
The $0_{th}$ order coefficient is set to zero, because the water resistance to a non-moving vessel is zero. 
Therefore, three of the parameters $\{p_1, p_2, p_{3}\}$ are the coefficients of the resistance curve. 
Typically the resistance curve is $2_{nd}$ degree polynomial. Here, a $3_{rd}$ order was used for improved expressivity.

\textbf{Rudder forces}: The rudder forces shown in~\cref{eq:forces} are associated with the rudder lift $c_L$ and drag $c_D$ coefficients.
These are modeled as 
$ c_L(a_r) = p_4 + p_5 a_r + p_6 a_r^2 + p_7 a_r^3 $ and $ c_D(a_r) = p_8 + p_9 a_r + p_{10} a_r^2 $, respectively. 
The polynomials degrees were selected to be as close as possible to their experimental measurements against rudder inflow angles $a_r$.
Lift and drag coefficients are the key factors involved in ship's turning abilities, hence seven $\{p_4, ..., p_{10}\}$ of the parameters are the polynomials coefficients of $c_L(\cdot)$ and $c_D(\cdot)$.

By targeting these specific elements, the model can be fine-tuned to capture ship-specific characteristics and operating conditions, enhancing its predictive accuracy while preserving the interpretability of the underlying physics-based model. The selected parameters affect the motion in all 3-DoF and capture typical sea trials, such as speed, turning and course instability tests. \vspace{-1mm}

\subsection{Parameter Optimization}
\label{subsec:param_opt}
To fit the parameters mentioned above to data of a specific vessel, we solve the following optimization problem, described by~\cref{eq:fit_opt}, where $M$ is the number of trajectories in the dataset. \vspace{-3mm}
\begin{equation}
\label{eq:fit_opt}
\begin{aligned}
& \underset{\mathbf{p}}{\text{min}}
& & \frac{1}{M} \sum_{i=0}^{M} \varepsilon(\xi_i, \bar{\xi}_i) \\
& ~ \text{s. t.}
& & \mathbf{b_l} \leq \mathbf{p} \leq \mathbf{b_u}, \\
& & & g(\mathbf{p}) \leq b_j, \; j = 1, \ldots, J
\end{aligned}
\end{equation}
This is a constraint nonlinear least squares (cNLS) problem, where $\varepsilon(\xi_i, \bar{\xi}_i)$ denotes the squared deviation between the predicted trajectory $\xi_i$ and the measured trajectory $\bar{\xi}_i$ over all timesteps. This is defined as \vspace{-1mm}
\begin{equation}
\label{eq:fit_cost}
    \varepsilon(\xi_i, \bar{\xi}_i) =  
    (\bar{\xi}_i - \xi_i)^T \mathbf{W} (\bar{\xi}_i - \xi_i) \vspace{-1mm}
\end{equation}
\noindent with $\mathbf{W} \in \mathbb{R}^{8\times8}$ being the diagonal weight matrix that balances the contribution of each dimension of the state.
$\mathbf{p} = \left[p_0, p_1, p_2, p_3, p_4, p_5, p_6, p_7, p_8, p_9, p_{10} \right]^T$ is the optimization variable, which is a vector of the key parameters described in~\cref{subsec:open_parameters}. $\mathbf{b_l}$ and $\mathbf{b_u}$ are the lower and upper bound vectors of parameters $\mathbf{p}$ that limit the range of possible values. $g(\mathbf{p})$ are non-linear functions of $\mathbf{p}$, which are constrained by $b_j$ to incorporate domain knowledge in the parameter optimization. The optimization problem is solved numerically using an interior-point gradient-based optimization method. This approach ensures that the estimated parameters minimize the weighted sum of squared deviations between measured and predicted trajectories, while adhering to the specified parameter bounds and additional constraints that prescribe the motion of the vessel.  \vspace{-1mm}

\section{Evaluation \& Results}
\label{sec:results}
\vspace{-1mm}

In this section, we validate the proposed parameter optimization approach and evaluate the resulting models in terms of their accuracy and reliability. 
The validation of our approach, involves fitting the data-driven physics-based model into data from two container ships (target ships), \textit{i.e.} given a dataset from one of the container ships the key parameters (see ~\cref{subsec:open_parameters}) of the data-driven physics-based model (see ~\cref{subsec:model}) are fitted by solving the cNLS optimization problem (see ~\cref{subsec:param_opt}). The evaluation of the fitted models involves comparing their accuracy and reliability against respective baseline physics-based models both numerically and visually. We consider two container ships of different sizes, containership A with 1750 DWT (approximately 85 meters in length) and the containership B with 8000 DWT (around 130 meters in length), to assess whether the parameter optimization process can: (a) converge to parameter values that are interpretable, physically meaningful, and (b) enable the fitted models predict motions across different ship configurations. Next, we describe the validation setup, including the details on: (i) the datasets used, (ii) the parameter optimization, (iii) the evaluation protocol, and (iv) the distance measures used. 


\textbf{Datasets}:
In our validation setup, we use synthetic datasets for both container ships. The maneuvers are based on realistic routes and speeds, focusing on port approaches and departures, with fewer open-sea scenarios. 
In-port maneuvers involve large drift angles which challenge most physics-based models. Fitting $\mathbf{p}$ to these maneuvers, we expect strong performance in simpler scenarios. Each dataset is generated using a distinct MMG-based simulator~\cite{yasukawa2015introduction,okuda2023maneuvering}, enabling evaluation of the optimization process's robustness to simulator-specific biases. These synthetic datasets, covering diverse vessel routes, allow comprehensive analysis under various conditions and support flexible maneuver testing.

The statistical summary of the datasets, shown in~\cref{table:errors}, reveals important characteristics that support the diversity of scenarios, \textit{e.g.} extensive range of $u$ and $n$. All trajectories have 120 knots and in terms of the datasets size (number of trajectories), for ship A, $M=165$ and for ship B, $M=100$, while the train-test split is conducted randomly to avoid any potential bias towards maneuvering or sea-keeping. The test scenarios for each vessel represent approximately the 12.5 \% of the trajectories. Next, we describe prominent aspects of the datasets.
\begin{table}[t]
\caption{Statistics of the datasets. The physical units of each dimension are as follows: $u$ and $v$ in $m/s$, $r$ in $rad/s$, $n$ is a real number, and $\delta$ in degrees. }
\centering
~~~~~~~~~~~~~~~~~~~\:\,
\begin{tabular}{|c||c|}
\hline
~~~~~~~~~~~~~~~~Train~~~~~~~~~~~~~~ & ~~~~~~~~~~~~~~Test~~~~~~~~~~~~~~~~\,\\
\end{tabular}\\
\parbox[t][4em][t]{4em}{\centering \vspace{-0.8cm} \textbf{Ship A \\ \vspace{1.2cm} Ship B}} 
\hspace{0.1cm} 
\begin{tabular}{|c|c|c|c|c|c||c|c|c|c|c|c|}
\hline
 & $u$ & $v$  & $r$  & $n$ & $\delta$  & $u$  & $v$  & $r$  & $n$ & $\delta$ \\ \hline
Min & -4.47 & -2.05 & -0.03 & -249.1 & -32.5 & 1.47 & -1.74 & -0.03 & -129.4 & -26.9 \\ \hline
Max & 8.23 & 1.76 & 0.03 & 249.4 &  33.5 & 7.38 & 1.76 & 0.03 & 245.9 & 27.9 \\ \hline
Mean & 3.27 & -0.02 & 0 & 95.0 &  1.2 & 4 & 0.09 & 0 & 129.1 & -0.04 \\ \hline
\multicolumn{11}{|c|}{\cellcolor{gray!10}{}} \\ 
\hline
Min & -0.47 & -1.04 & -0.02 & -158.8 & -40.0 & 0.1 & -0.57 & -0.01 & -98.9 & -39.5 \\ \hline
Max & 6.17 & 1.03 & 0.02 & 169.7 & 40.0 & 5.45 & 0.54 & 0.01 & 168.8 & 32.6 \\ \hline
Mean & 3.83 & 0.04 & 0 & 93.3 & -1.0 & 3.71 & 0.11 & 0 & 88.8 & -0.8  \\ \hline
\end{tabular}
\label{table:errors}
\vspace{-4mm}
\end{table}

 

\noindent
\underline{Speed Characteristics}: 
As we can see in~\cref{table:errors} the datasets cover a large portion of the operational conditions the vessels may encounter during their lifetime. 

\noindent
\underline{Maneuvering Characteristics}: 
Sway speeds, and yaw rate of turn indicates inclusion of broad maneuvering behaviors. 

\noindent
\underline{Propulsion Characteristics}: The range $n$ differs between vessels, yet it is consistent between the respective training and test sets. Although the vessels have different propulsion system characteristics (different MCR) the range of $n$ indicates that almost all operational conditions are present in the datasets.



In summary, the differences within each ship's data and across vessels create an ideal validation setup to assess the approach's ability to capture vessel-specific behaviors that generalize.

\textbf{Parameter Optimization}:
For the cost function of~\cref{eq:fit_cost}, we select the hyperparameter $\mathbf{W} = diag(\frac{1}{\bar{L}_i}, \frac{1}{\bar{L}_i}, \frac{1}{\pi}, \frac{1}{\bar{U}_i}, \frac{1}{\bar{U}_i}, \frac{1}{max(r_i)}, 0, 0)$ as we  aim to minimize the discrepancies of the predicted trajectory $\xi_i$ and the ground-truth trajectory $\bar{\xi}_i$ in terms of pose and velocities dimensions. $n$ and $\delta$ are assumed to be identical in $\xi_i$ and $\bar{\xi}_i$. The denominators in $\mathbf{W}$ rescale and nondimensionalize the different dimensions. $\bar{L}_i$ is the Cartesian length of $\bar{\xi}_i$, $\bar{U}_i$ is the average speed of $\bar{\xi}_i$ and $max(r_i)$ the maximum angular velocity set to be 0.0314 rad/s and the 
initial guess of parameter vector for ship A is $[34187.03, -12569.98, 1586.29, 0.00, $ $ -0.02, 2.70, 0.42, -3.23, 0.06, -0.11, 1.97]^T$ and for ship B is $[5504.65, -4218.06, $ $ 1063.42, -0.05, -0.01, 2.73, 0.35, -3.10, 0.06, -0.11, 1.97]^T$.

One could solve~\cref{eq:fit_opt} without using any constraints (non-linear least squares), yet parameters' physical consistency may be compromised, making the fitted model less interpretable.
Thus, we incorporate constraints $g(\cdot)$ that are functions of $\mathbf{p}$ and bounds $b_j$ both derived from marine engineering domain expertise. 
These are: (i) $0 \leq R(u) \leq 80000 $, (ii) $ \dot{R}(u) > 0 $, (iii) $ -0.05 X_{\delta} \leq Y_{\delta} \leq 0.05 X_{\delta}$, (iv) $ -1.0 \leq c_L(a_r) \leq 0, \forall a_r < 0$ and $ 0 \leq c_L(a_r) \leq 1.0, \forall a_r \geq 0$, (v) $ 0 \leq c_D(a_r) \leq 1.5$ and $c_D(0) \leq 0.05$,
while all elements of $\mathbf{b_l}$ are set to \texttt{-inf} and for $\mathbf{b_u}$ to \texttt{inf}.  

\textbf{Evaluation Protocol}:
We compare three distinct trajectories for each ship: stemming from ground truth $\bar{\xi}_i$, baseline physics-based $\tilde{\xi}_i$, and our fitted model $\xi_i^*$. To do that, for each ship we quantify the distance between
(i) the ground truth and the baseline trajectories, (ii) the ground truth and fitted model trajectory, and report the relative improvement from (i) to (ii).

\textbf{Distance Measures}:
To quantify the distance between two trajectories, we use a Manhattan Distance (MD) per dimension,
and a custom Vessel Distance Measure (cVDM) that rescales and nondimensionalizes the different dimensions, both given in~\cref{eqref:MD,eqref:cVDM}. Notation-wise see~\cref{sec:framework}. For a study on different measures and their utility see~\cite{mathioudakis2025realworldvalidationphysicsbasedship}, where we show the aptness of cVDM (over other measures, such as MD) to quantify deviations between trajectories. Also, note that given the selected $\mathbf{W}$ discussed above, \cref{eq:fit_cost} is the differentiable equivalent to cVDM.  To evaluate~\cref{eqref:MD,eqref:cVDM} for $\tilde{\xi}_i$ and $\xi_i^*$, we replace the variables accordingly, \textit{e.g.} $\tilde{\xi}_i$, $\mathbf{s}_{d,k}$ with $\tilde{\mathbf{s}}_{d,k}$, and $\xi_i^*$, $\mathbf{s}_{d,k}$ with $\mathbf{s}^*_{d,k}$.
\vspace{-2mm}
\begin{align}
\label{eqref:MD}
     & \hspace{3cm} MD = \frac{1}{K} \sum_{k=1}^{K} \left|\bar{\mathbf{s}}_{d,k} - \mathbf{s}_{d,k} \right|
\end{align}
\vspace{-3mm}
\begin{align}
\label{eqref:cVDM}
    cVDM &= \frac{100}{K} \sum_{k=1}^{K} \bigg(
        \frac{\left| \bar{x}_{k} - x_{k} \right|}{\bar{L}}
        + \frac{\left| \bar{y}_{k} - y_{k} \right|}{\bar{L}} + \frac{\left| \bar{\psi}_{k} - \psi_{k} \right|}{\pi} + \notag \\
        &\quad \quad \quad \quad \quad ~
        \frac{\left| \bar{u}_{k} - u_{k} \right|}{\bar{U}}
        + \frac{\left| \bar{v}_{k} - v_{k} \right|}{\bar{U}}
        + \frac{\left| \bar{r}_{k} - r_{k} \right|}{r_{max}} \bigg)
\end{align}
\vspace{-5mm}


\subsection*{Results:}
\vspace{-1mm}


Here, we report the fitted parameters for both target ships based on the respective datasets and discuss their interpretation. Next, we present the overall performance following the evaluation protocol and inspect qualitatively and quantitatively individual scenarios.     
 
\begin{figure}[t]
    \centering
    \includegraphics[width=0.99\linewidth]{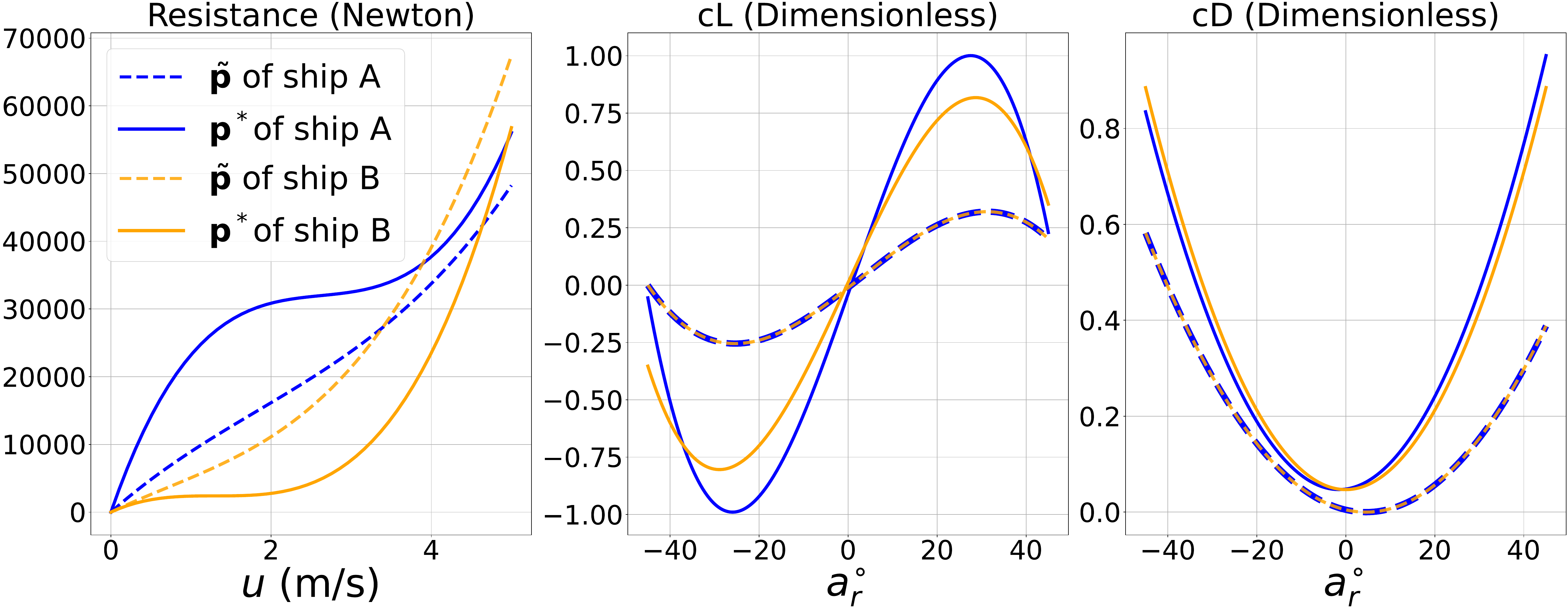}
    \vspace{-2mm}
    \caption{Fitted ($\mathbf{p^*}$) vs baseline ($\mathbf{\tilde{p}}$) model curves for resistance, lift ($c_L$), and drag ($c_D$) coefficients for both vessels. Note that both ships have the same rudder type.} 
    \label{fig:fitted_curves}
    \vspace{-4mm}
\end{figure}

\textbf{Fitted parameters}: 
\Cref{table:coeffs} includes the parameters used for the baseline models of each ship and the ones of the fitted models, denoted with $\mathbf{\tilde{p}}$ and $\mathbf{p^*}$, respectively. $\mathbf{\tilde{p}}$ are calculated based on captive test data. To inspect the different sets of parameters, we plot (see~\cref{fig:fitted_curves}) six curves for each ship, two for $R(u)$, two for $c_L$ and two for $c_D$. All curves follow marine engineers expectations, yet they are significantly different to the ones of the baselines. $c_L$ demonstrates a S-shape and $c_D$ a parabolic pattern, while both maintain the asymmetry which is also observed in the baseline curves. The curves suggest that the fitted models still obey the underlying physics, hence the models remain interpretable. 

\textbf{Overall performance}: 
We obtain a relative improvement based on MD, referred to as $mARI$\footnote{For 6 dimensions and $M$ trajectories $MD \in \mathrm{R}^{6 \times M}$. Average relative improvement for each trajectory 
is the set $\{ \alpha_0, ..., \alpha_M \}$ and $mARI$ is the median of the set.} and based on $cVDM$. For ship A, $mARI$ is 32.1\% and $cVDM$ is 51.6\%, while for ship B $mARI$ is 42.9\% and $cVDM$ is 57.8\%. These percentages indicate how much closer to the groundtruth are the trajectories of the fitted models compared to the baselines, while the predictions are 72.36\% 
\begin{table}[!h]
\vspace{-2mm}
\caption{Parameters for both vessels. $\mathbf{\tilde{p}}$ denotes parameters of the baseline models while $\mathbf{p^*}$ denotes parameters of the fitted models.}
\vspace{-1mm}
\centering
~~~~~~~~\:
\begin{tabular}{|c|c|c|c|}
\hline
 ~~$Y_n$~~\, & ~~~~~~~~$R(u)$~~~~~~~~~\: & ~~~~~~~~~~~$c_L(a_r)$~~~~~~~~~~~~~~ & ~~~~~~~~$c_D(a_r)$~~~~~~~~~\\
\end{tabular}\\
\parbox[t][4em][t]{0em}{\centering \textbf{\vspace{-1.4cm} \footnotesize{Ship} \\ \vspace{0.2cm} ~~ A \\ \vspace{0.9cm} ~~ B}} 
\hspace{0.6cm} 
\begin{tabular}{|c|c|c|c|c|c|c|c|c|c|c|c|}
\hline
  & $p_0$ & $p_1$ & $p_2$ & $p_3$ & $p_4$ & $p_5$ & $p_6$ & $p_7$ & $p_8$ & $p_9$ & $p_{10}$ \\ \hline
$\mathbf{\tilde{p}}$ & 0.000 & 10500 & -1900  & 346 & -0.012 & 0.864 & 0.182 & -1.191 & 0.005 & -0.1230 & 0.779 \\ \hline
$\mathbf{p^*}$  & -0.017 & 34187 & -12568 & 1594 & -0.039 & 3.193 & 0.205 & -4.882 & 0.048 & 0.0746 & 1.370 \\ \hline
\multicolumn{12}{|c|}{\cellcolor{gray!10}{}} \\ \hline
$\mathbf{\tilde{p}}$  & 0.000 & 5669 & -1127 & 538 & -0.012 & 0.864 & 0.182 & -1.191 & 0.005 & -0.1230 & 0.779 \\ \hline
$\mathbf{p^*}$ & -0.027 & 5505 & -4215 & 1077 & 0.012 & 2.420 & -0.019 & -3.195 & 0.047 & 0.0001 & 1.359 \\ \hline
\end{tabular}
\label{table:coeffs}
\vspace{-1mm}
\end{table}
and 89.67\% more consistent\footnote{The variance of $cVDM$ of the fitted model is x\% lower than the one of the baseline.}, for ships A and B, respectively.

\begin{figure}[h]
    \centering
    \includegraphics[width=0.99\linewidth]{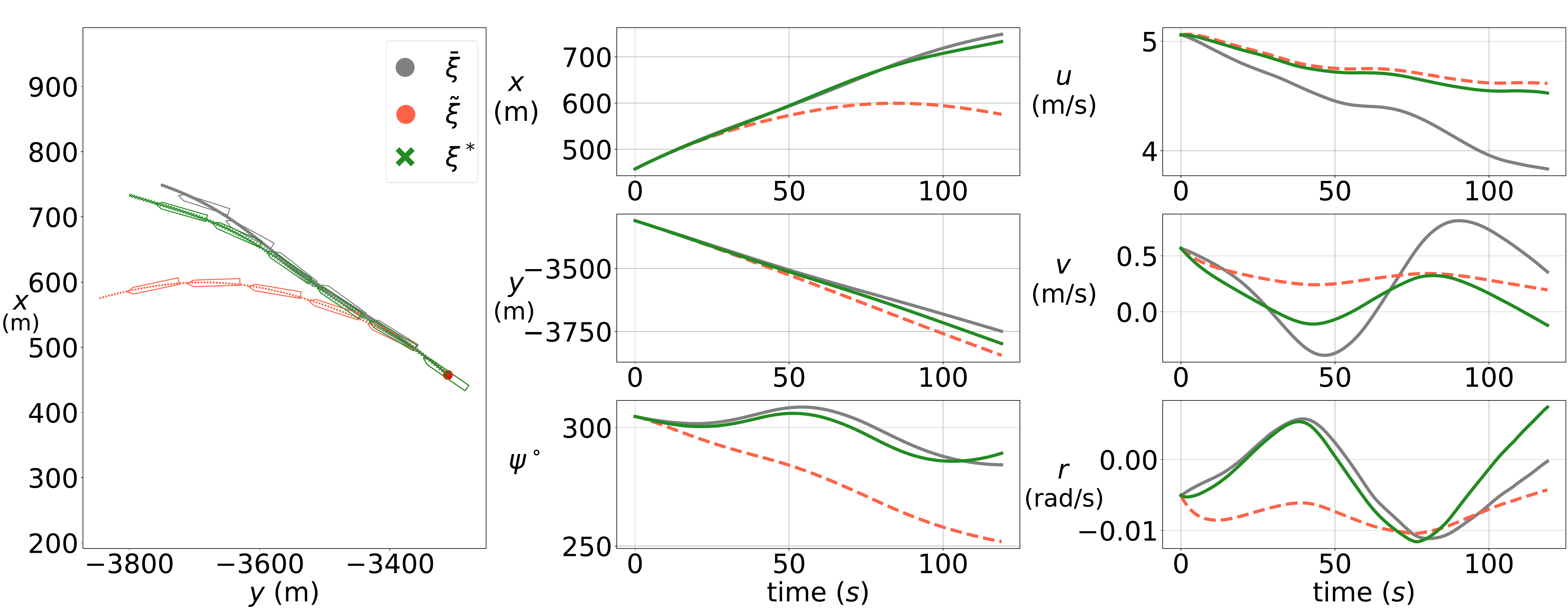}
     \vspace{-2mm}
    \caption{Trajectory comparison for ship A with excellent accuracy improvement. Generated trajectory: $\bar{\xi}$ from the groundtruth data, $\tilde{\xi}$ from the baseline model and $\xi^*$ from the fitted model.   MD$(\%)$ per dimension is: $x:90.6$, $y:56.4$, $\psi:86.9$, $u:13.0$, $v:13.1$, $r:52.2$, and cVDM$(\%)$ is 76.0.}
    \label{fig:Ship_A_good}
    \vspace{-3mm}
\end{figure}

\begin{figure}[t]
    \centering
    \includegraphics[width=0.99\linewidth]{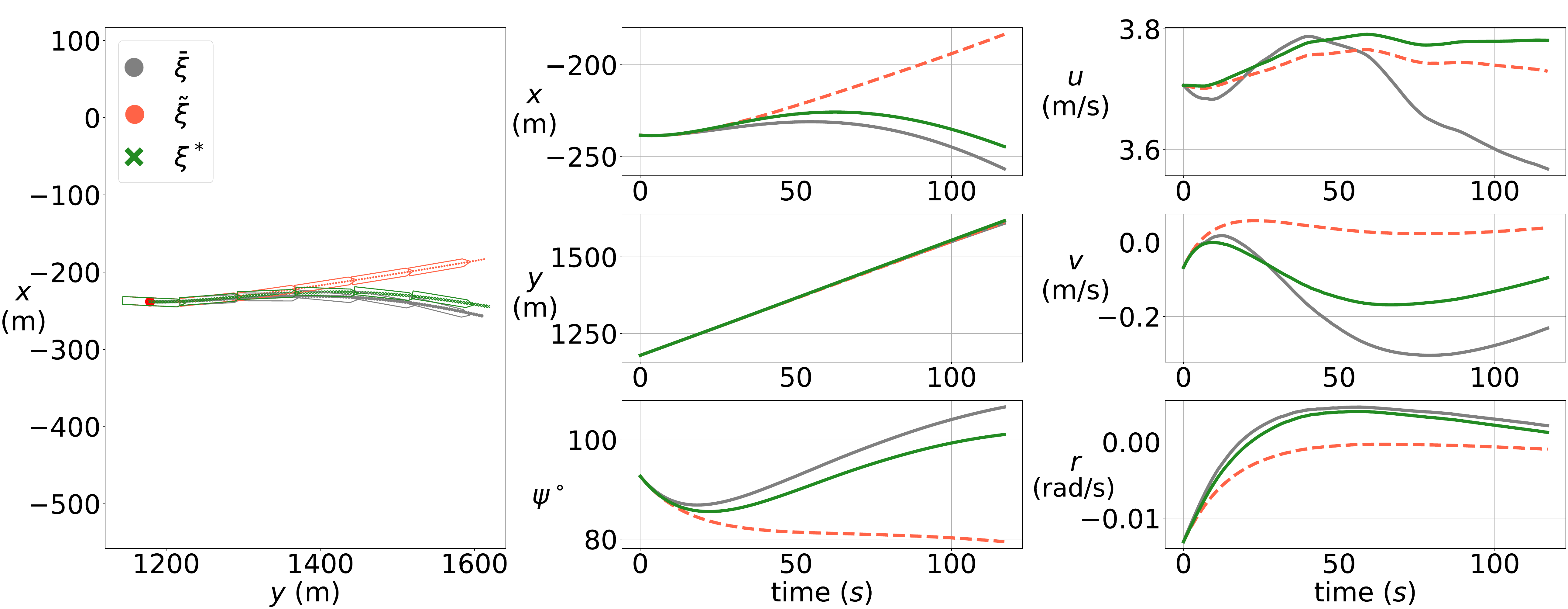}    \vspace{-2mm}
    \caption{Trajectory comparison for ship B with remarkable  accuracy improvement. Generated trajectories same as in~\cref{fig:Ship_A_good}.   MD$(\%)$ per dimension is: $x:75.7$, $y:-80.3$, $\psi:77.3$, $u:-25.9$, $v:61.2$, $r:80.9$, and cVDM$(\%)$ is 69.2.} 
    \label{fig:Ship_B_good}
    \vspace{-2mm}
\end{figure}

\begin{figure}[t]
    \centering
    \includegraphics[width=0.99\linewidth]{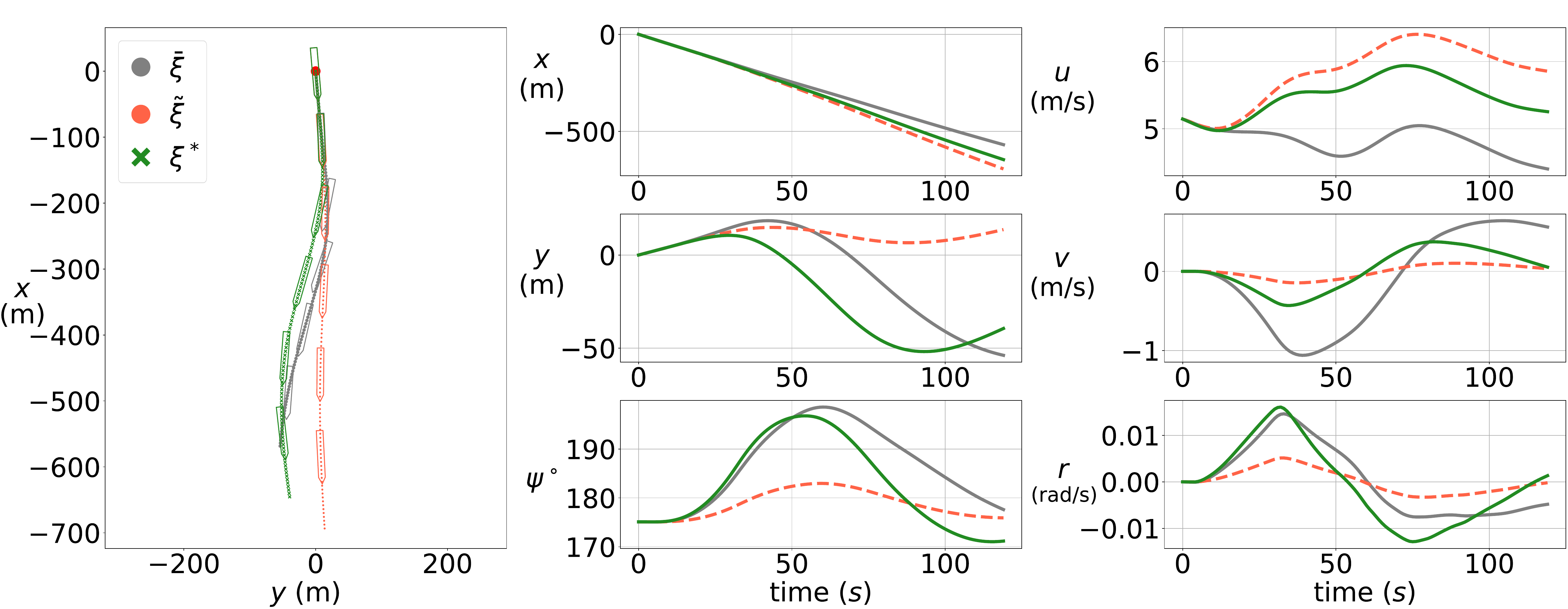}     \vspace{-2mm}
    \caption{Trajectory comparison for ship A with increased accuracy and minor overshoot. Generated trajectories same as in~\cref{fig:Ship_A_good}.  MD$(\%)$ per dimension is: $x:36.3$, $y:22.1$, $\psi:43.9$, $u:34.5$, $v:23.5$, $r:31.9$, and cVDM$(\%)$ is 28.0.} 
    \label{fig:Ship_A_moderate}
    \vspace{-2mm}
\end{figure}

\begin{figure}[t]
    \centering
    \includegraphics[width=0.99\linewidth]{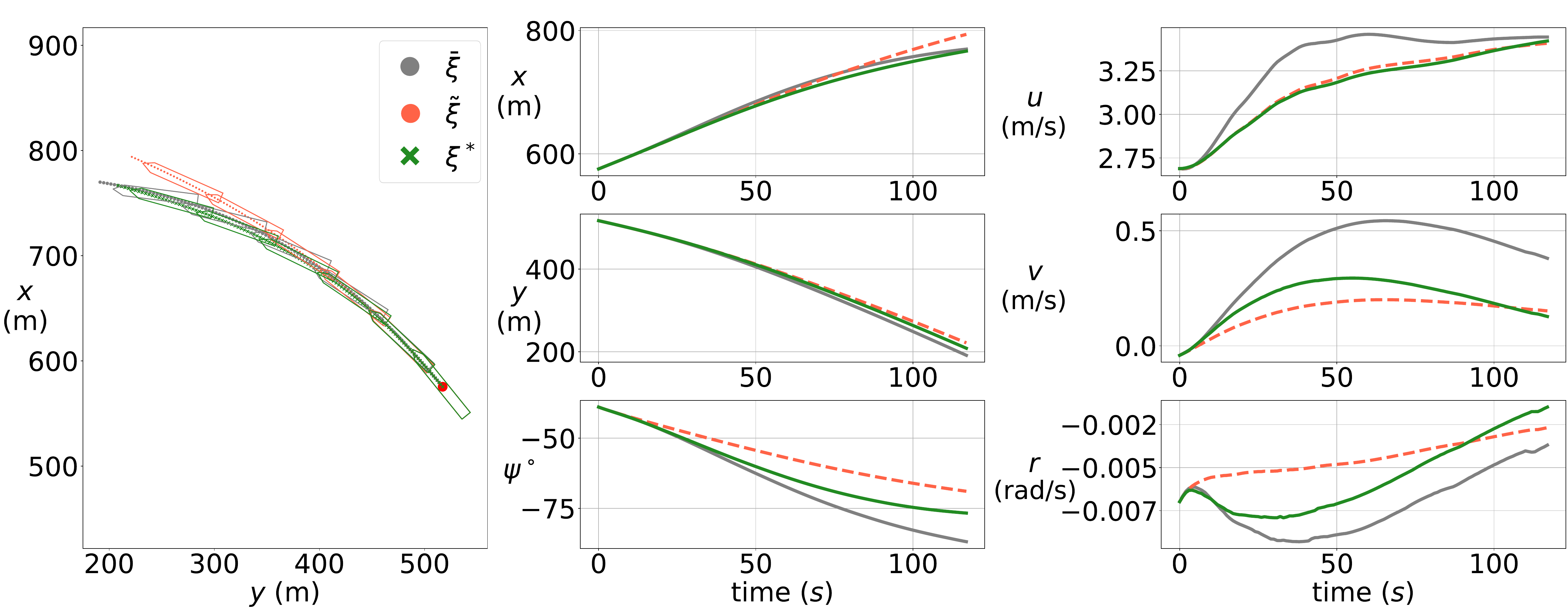}     \vspace{-2mm}
    \caption{Improved trajectory prediction for ship B with overreaction (i.e. in $x$ and $u$). Generated trajectories same as in~\cref{fig:Ship_A_good}. MD$(\%)$ per dimension is: $x:-6.7$, $y:36.1$, $\psi:58.8$, $u:-9.4$, $v:22.3$, $r:41.7$, and cVDM$(\%)$ is 22.5.} 
    \label{fig:Ship_B_moderate}
    \vspace{-3mm}
\end{figure}

\textbf{Distinct maneuver scenarios}:
Here, we discuss four distinct maneuver scenarios (two from ship A and two from ship B). Please note that the chosen test trajectories presented here are not just the most optimal but the most instructive ones. First category (see~\cref{fig:Ship_A_good,fig:Ship_B_good}) of scenarios show remarkable improvement of cVDM = 76.0\% for ship A and cVDM = 69.2 \% for ship B, which is also verified by visual inspection. Second category (see~\cref{fig:Ship_A_moderate,fig:Ship_B_moderate}) of scenarios where the fitted models shows moderate improvement over the baseline. Ship A has cVDM = 28.0\% and ship B has cVDM = 22.5\%. The latter, indicate existence of unmodelled phenomena that we plan to study in future work. 

\textbf{Cross-vessel takeaways}:
The fitted models' performance across both vessels reveals several key insights: (i) maintain physical consistency and interpretability, 
(ii) consistent improvement in terms of prediction accuracy and consistency over the baselines regardless of the ship, and (iii) reliable performance across different operating conditions and speed ranges. \vspace{-2mm}


\section{Conclusion and future work}
\label{sec:conclusion}


In this work we introduce a novel data-driven physics-based modeling approach that successfully combines physics-based models with data-driven techniques for marine vessel trajectory prediction. We present the core physics-based model, its key parameters and a cNLS parameter optimization method. The  validation across two distinct vessels, ship A and ship B, demonstrates: (i) substantial and consistent improvement in prediction accuracy across different operational regimes, and (ii) successful adaptation to vessel-specific characteristics while maintaining physical interpretability  of the fitted models. By utilizing the same data-driven adaptation process for both ships, we showcase the versatility and applicability of the approach in handling different vessel characteristics. In terms of future work, we plan to investigate additional key parameters relevant to environmental factors (wind, waves, currents) to enhance the adaptability of the data-driven models in broader range of conditions. \vspace{-2mm}

\section{Acknowledgments}
\vspace{-2mm}

This research is part of the developments with Nabtesco corporation for the Nippon Foundation \texttt{MEGURI2040} Fully Autonomous Ship Program. \vspace{-2mm}

\bibliographystyle{splncs04}
\bibliography{references}

\end{document}